\documentclass[]{spie}  

\usepackage{graphicx}

\usepackage{xspace}
\usepackage{siunitx}
\usepackage{amssymb,amsmath,amsfonts}
\usepackage[colorlinks=true, allcolors=blue]{hyperref}
\usepackage{booktabs}
\usepackage{ulem}
\usepackage[T1]{fontenc}

\usepackage{upgreek}

\newcommand{\superbit}{\textsc{SuperBIT}\xspace}

\title{Overview, design, and flight results from SuperBIT: a high-resolution, wide-field, visible-to-near-UV balloon-borne astronomical telescope}

\author[a,b,c,d]{L. Javier Romualdez}
\author[e]{Steven J. Benton}
\author[d]{Anthony M. Brown}
\author[d]{Paul Clark}
\author[a]{Christopher J. Damaren}
\author[f,g]{Tim Eifler}
\author[e]{Aurelien A. Fraisse}
\author[b,c]{Mathew N. Galloway}
\author[b,c]{John W. Hartley}
\author[h]{Mathilde Jauzac}
\author[e]{William C. Jones}
\author[e]{Lun Li}
\author[e]{Thuy Vy T. Luu}
\author[d]{Richard J. Massey}
\author[f]{Jacqueline Mccleary}
\author[b,c,i,j]{C. Barth Netterfield}
\author[a,c]{Susan Redmond}
\author[f,k]{Jason D. Rhodes}
\author[d]{J\"urgen Schmoll}
\author[h]{Sut-Ieng Tam}

\affil[a]{University of Toronto Institute for Aerospace Studies (UTIAS), 4925 Dufferin Street, Toronto, ON, Canada}
\affil[b]{Department of Physics, University of Toronto, 60 St. George Street, Toronto, ON, Canada}
\affil[c]{Dunlap Institute, University of Toronto, 50 St. George Street, Toronto, ON, Canada}
\affil[d]{Centre for Advanced Instrumentation (CfAI), Durham University, Science Laboratories, South Road, Durham, UK}
\affil[e]{Department of Physics, Princeton University, Washington Road, Princeton, NJ, USA}
\affil[f]{Jet Propulsion Laboratory (JPL), California Institute of Technology, 4800 Oak Grove Drive, Pasadena, CA, USA}
\affil[g]{Department of Astronomy/Steward Observatory, 933 North Cherry Avenue, Tucson, AZ 85721-0065, USA}
\affil[h]{Centre for Extragalactic Astrophysics, Durham University, South Road, Durham, UK}
\affil[i]{Department of Astronomy, University of Toronto, 50 St. George Street, Toronto, ON, Canada}
\affil[j]{Canadian Institute for Advanced Research, 661 University Ave., Suite 505, Toronto, ON, Canada}
\affil[k]{California Institute of Technology, 1201 East California Blvd, Pasadena, CA, USA}

\authorinfo{Further author information: (Send correspondence to J. Romualdez)\\ Address: Department of Physics University of Toronto, 60 St. George St., Toronto, ON, Canada M5S 1A7\\ E-mail: javier.romualdez@mail.utoronto.ca, Telephone: 1-416-946-0946}

\begin{document} 
\maketitle

\begin{abstract}
Balloon-borne astronomy is a unique tool that allows for a level of image stability and significantly reduced atmospheric interference without the often prohibitive cost and long development time-scale that are characteristic of space-borne facility-class instruments. The Super-pressure Balloon-borne Imaging Telescope (SuperBIT) is a wide-field imager designed to provide 0.02" image stability over a 0.5 degree field-of-view for deep exposures within the visible-to-near-UV (300-900 um). As such, SuperBIT is a suitable platform for a wide range of balloon-borne observations, including solar and extrasolar planetary spectroscopy as well as resolved stellar populations and distant galaxies. We report on the overall payload design and instrumentation methodologies for SuperBIT as well as telescope and image stability results from two test flights. Prospects for the SuperBIT project are outlined with an emphasis on the development of a fully operational, three-month science flight from New Zealand in 2020.
\end{abstract}


\section{INTRODUCTION}
\label{sec:intro}

\subsection{Astronomical \& Cosmological Background}

The concept of an inexpensive, near-space quality observing platform with instruments that can be replaced and easily upgraded has enormous potential for scientific discovery. 
While still in its early stages, NASA's recently developed super pressure balloon (SPB) capability can provide the means to enable such a platform from stratospheric altitudes, especially in combination with high-resolution, wide-field imagers or spectrographs. 
Altogether, the combination of diffraction limited angular resolution, extreme stability, space-like backgrounds, and long integrations enable transformative opportunities for astrophysics and cosmology. 

Numerous modern science objectives require these observing conditions, which, over the past couple of decades, has driven a highly oversubscribed demand for UV-visible-NIR observations on the Hubble Space Telescope (HST). 
While JWST (James Web Space Telescope), Euclid, and WFIRST (Wide-Field Infrared Survey Telescope) will surpass HST's capabilities at red and NIR (Near Infrared) wavelengths, the inevitable demise of HST will result in effectively no space-based capabilities in the blue and UV. 
Even in the visible and NIR, planned missions will not begin to exhaust the demand for wide-field, high-resolution imaging. 
In this way, SPB based imaging and spectroscopy has enormous potential to advance a large fraction of research topics in astronomy and cosmology.
These include:

\begin{enumerate}
\item Exoplanets: searches and classification via imaging and spectroscopy.
\item Interstellar Physics: large-scale mapping of the interstellar medium;
resolved imaging and spectroscopy of star forming regions; evolution of molecular abundances.
\item Stars: Young-star accretion variability; simultaneous star observations in the UV and NIR; astroseismology; search for pulsars in local group; search for white dwarfs in galactic halo.
\item Galaxies: Modified gravity constraints from galaxy morphology observations; galaxy evolution, e.g. a wide-field COSMOS survey; line-intensity mapping; study merger history and star formation triggers; study high-z galaxy morphology.
\item Black Holes: Low-mass AGN reverberation mapping in the UV-optical-NIR; variability of AGN; Gamma Ray Burst monitoring; black hole accretion disk dynamics; high-resolution imaging of blazars (highly compact, energetic radio sources).
\item Planetary Science: Comet water studies; UV comet imaging; wide-field survey for NEOs (Near Earth Objects); internal structure from Jupiter and Saturn via surface oscillations, study solar system moons, spectroscopy of planet atmospheres; deep imaging of Kuiper Belt Objects. 
\item Dark Energy/Dark Matter: Dark matter mapping of Active Galactic Nuclei (AGN); Strong Lensing time-delays; wide-field weak lensing to calibrate ground based shape measurements; combined imaging and spectroscopic weak lensing measurements.
\item Galaxy Clusters: galaxy cluster lensing mass surveys; supernovae observations; precision UV/visual photometric measurements complementary to future ground and space based missions.
\end{enumerate}

\noindent In the following we discuss briefly some of the observationally-motivated details of the last two science topics. 
Dark Energy and Dark Matter together constitute 96\% of the content of the Universe's energy density, yet their nature remains poorly understood. 
To this end, wide-field astronomical surveys during the next decade have potential to explore both concepts, via statistical measurements of e.g.\ large-scale structure, gravitational lensing, and redshift space distortions. \cite{MasseyLensing}
Amongst planned facilities, the ground-based Large Synoptic Survey Telescope (LSST)\cite{LSST} will explore a large volume of the Universe (18,000 deg$^2$ to 27.5 r-band magnitude), but will be limited by atmospheric seeing. 
Future space-based missions will have exquisite image quality but will be either substantially shallower or cover a smaller fraction of the sky. 
Euclid\cite{Euclid}, which will overlap with LSST for $\sim$6000 deg$^2$, will be $\sim 2$ magnitudes shallower, whereas WFIRST\cite{WFIRST} will be as deep as LSST, but its imaging will cover only 2300 deg$^2$.  

High-resolution observations from a stable platform are especially important to control systematic errors in cosmological probes based on weak gravitational lensing (WL): e.g.\ cosmic shear, galaxy-galaxy lensing, cluster/filament weak lensing, void lensing. 
WL methods rely on accurately measuring the shapes and redshifts of large ensembles of galaxies, most of which are relatively small and faint.
Controlling uncertainties in shape and redshift measurement is difficult from the ground, due to atmospheric distortions and limitations in wavelength coverage, in particular towards the blue/UV. 
More concretely, the benefit that a potential Euclid, LSST, and WFIRST joint data set would gain from a SPB platform is twofold. 
First, none of these missions offers the capability for space-quality resolution blue/UV imaging.
Second, Euclid's wide-field survey is too shallow to calibrate shape measurement for LSST's faintest galaxies, and WFIRST's measurements are limited to the IR, where intrinsic morphologies are different. 
A deep large-area SPB survey could fill the mismatches between these surveys, substantially reducing systematics
Similarly, a deep targeted SPB survey could follow up regions of interest, such as merging `bullet' clusters.

\begin{figure}
	\centering
	\begin{tabular}{@{}cc@{}}
	\includegraphics[width=0.5\textwidth]{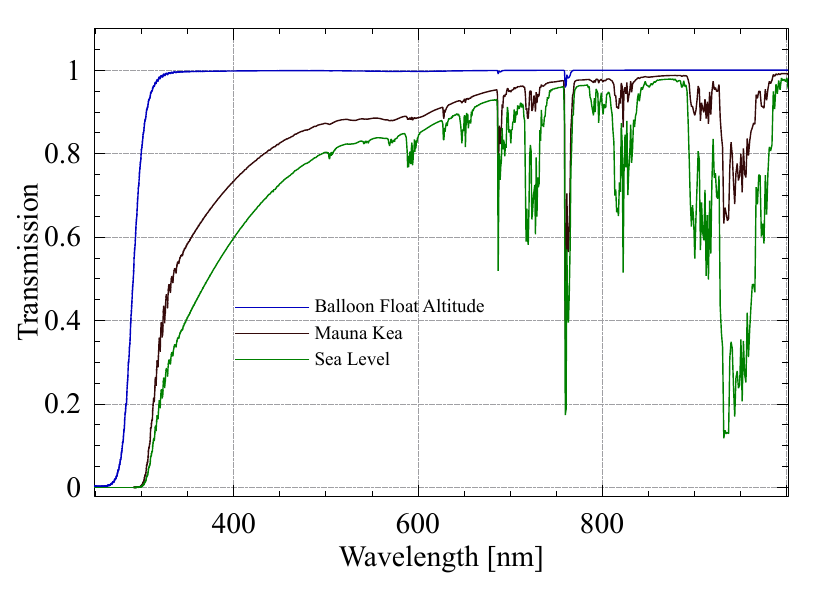} & \includegraphics[width=0.5\textwidth]{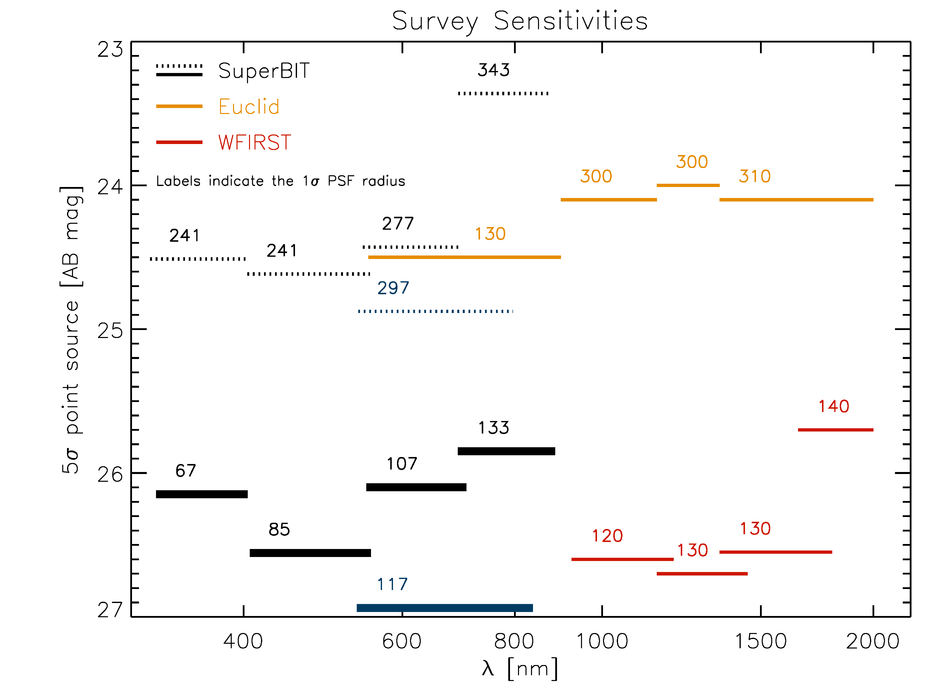}
	\end{tabular}
	\caption{(Left) atmospheric transmission at various altitudes as calculated by the MODTRAN4 software; there is significantly reduced atmospheric absorption in the visible-to-near-UV, notably for wavelengths below 400 nm, compared to even the best ground-based observing sites. (Right) Survey sensitivities of various upcoming observatories for a $5\sigma$ point-source in the visible-to-near-UV spectrum}
	\label{fig:balloontrans}
	\end{figure}
    
\subsection{High Precision Balloon-borne Astronomy}
At an altitude of 35-40 km, mid-latitude ($\sim 45^\circ$ S) balloon-based observations offer the unique opportunity to undertake targeted research in the aforementioned areas of cosmology while directly contributing to multiwavelength surveys from ground- or space-based observations to enhance their science return. 
Given the capability for SPB platforms for high payload mass ($\sim$1200 kg) and reduced power constraints, sub-orbital missions of this kind benefit from dramatically reduced development, launch, and operational costs compared to space-borne instruments, which allows for greater institutional accessibility during a typical 70--100 day observational run.
Coupled with the overall recoverability of balloon-borne payloads, which allows for quick turnaround for potentially annual instrument redeployment, SPB platforms can afford both late-development technology freeze as well as the ability to qualify prospective space-borne instrumentation, such as, for example, cutting-edge detector technology and lightweight, high fidelity optical elements.
As a result of these clear advantages, the science return of multiple balloon flights can rival or exceed that of high-profile UV-visible space missions at around 1\% of their cost. 
Although, at the time of this publication, SPB capabilities are still in the development stages, two SPB missions have successfully flown from Wanaka, New Zealand, with plans to further qualify the technology with a science-capable mission by 2020.

Despite the clear advantages promised and demonstrated by the SPB launch platform, scientific ballooning in general presents its own unique challenges that are not typically encountered by traditional ground- or space-based approaches.
Structurally, a typical stratospheric launch vehicle consists of a 1 million cubic meter helium balloon tethered to a 80--100 m long \textit{flight train} containing the parachute and attached to the scientific payload (or \textit{gondola}) through an actuated pivot connection (see Fig. \ref{fig:BIT_profile}).
From a dynamical perspective, the induced and inevitably persistent compound pendulations that manifest as a result dominate the low-frequency regime with $\sim 0.05$ Hz pendulations about the balloon connection and $\sim 1$ Hz pendulations about the pivot connection at around the 1-2 arcminute level.
In addition to torsional and modal effects from the flight train driven at low-to-mid frequencies of 0.1-3 Hz by stratospheric wind shears and balloon dynamics, any flexible structural modes in the scientific payload are amplified at the 1-30 arcsecond level, which can be detrimental optically-based observations. 
Thermally, day-night cycles at mid-latitudes within a radiation-dominated environment (stratospheric pressure at $\sim 3$ mbar) gives rise to potential difficulties with maintaining alignment of sensitive optical components, where thermal gradients across the gondola can be detrimental to overall optical performance and stability. 

Outlined in this paper is an overview of the approaches and methodologies used to mitigate these sub-orbital effects in a way that provides a level of structural and optical stability sufficient for cosmological research from balloon-borne payloads.
Techniques for three-axis telescope stability as well as further high-bandwidth image stabilization within a harsh sub-orbital environment are summarized with results from engineering flights, which demonstrates the viability of scientific balloon-based astronomical observations for wide-field, deep imaging in the visible-to-near-UV spectrum.
A rigorous treatment of balloon-borne dynamic stability and an in-depth analysis of flight results for these particular engineering approaches is available in literature.\cite{RomualdezThesis18}

\section{BALLOON-BORNE IMAGING TELESCOPE (SUPERBIT)}
\label{sec:superbit}
The Super-pressure Balloon-borne Imaging Telescope, also known as \superbit, is an astronomical instrument designed to demonstrate the capabilities and performance of high-gain stability, on-demand, wide-field, visible-to-near-UV imaging from the stratosphere as a cost-effective, robust, and reusable platform for strong/weak lensing experiments, exo-planetary studies, and other cosmological studies. 
\superbit has been developed and is currently designed to take advantage of the primary benefits of balloon-borne optical observations, namely pristine atmospheric transmission at stratospheric altitudes, nearly negligible cost compared to space-borne instruments of a similar class, and potential for quick science turnaround by utilizing SPB capabilities.
From an instrumentation perspective, \superbit, at the time of this publication, houses a 0.5 m Ritchey-Cretien-type telescope with refractive field correction optics providing a relatively wide $0.5^\circ$ field-of-view, which is stabilized in all three rotational axes. 
To provide diffraction-limited imaging within the dynamically harsh balloon-borne environment, \superbit has the precision, accuracy, and pointing fidelity to acquire science targets to within < $0.5^\prime$, to stabilize the $\sim 80$ kg telescope payload to sub-arcsecond levels, and to further mitigate sub-arcsecond disturbances on the telescope focal plane to within a $0.02^{\prime\prime}$ image stability (1-$\sigma$) over integration runs on the order of an hour. 
A brief description of the mechanical architecture and control approach used for the \superbit instrument is provided in the following sections.

\subsection{Mechanical Architecture \& Stabilization Hardware}
\begin{figure}
\centering
\includegraphics[width=0.8\textwidth]{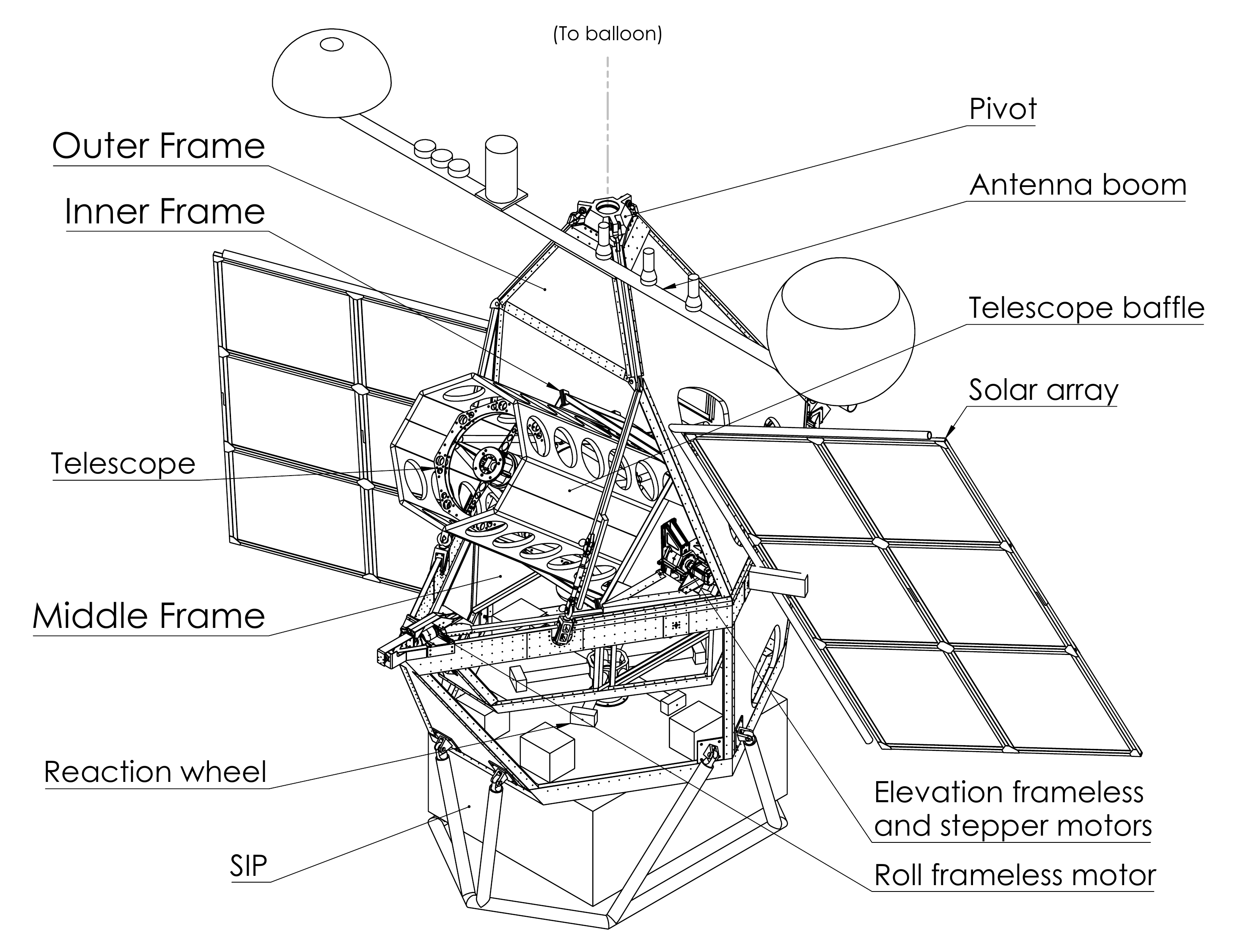}
\caption{The \superbit gondola as of the Palestine, 2018 campaign; the main structure is comprised of three independently rotating gimbals, constructed primarily out of aluminum honeycomb panels; during flight, the gondola communicates with the ground through the support instrumentation package (SIP), which is a modular isolated flight system given by the launch provider}
\label{fig:BIT_profile}
\end{figure}

Mechanically, \superbit is a three-axis stabilizer that compensates for the rough compound pendulative effects that manifest in the balloon-borne environment while compensating for field rotation effects over a relatively wide $0.5^\circ$ field-of-view.
When fully assembled, the \superbit instrument is approximately 3 m tall from the base to the flight train connection or \textit{pivot} and weighs approximately 1000 kg, which is within the weight class suitable for SPB missions.
Referring to \ref{fig:BIT_profile}, the three mechanical degrees-of-freedom are constructed from a series of nested frames: an outer frame, which controls gondola yaw; a middle frame, which actuates gondola roll about the outer frame; and an inner frame, which controls the telescope pitch about middle frame.
Constructed primarily from aluminum honeycomb panels, these three nested frames are both lightweight and structurally stiff, which reduces the potential for the driving of structural modes that can degrade pointing performance and subsequent image quality. 
While the full $360^\circ$ yaw range is accessible from altitude, the roll of the middle frame is constrained to $\pm 6^\circ$ due to inter-frame interference and the pitch of the inner frame is restricted at $20^\circ$ on the lower end due to the Earth's limb and $55^\circ$ on the upper end from the angular obstruction caused by the helium balloon at full expansion.
A complete description of the structural design for \superbit is provided in literature\cite{StevenThesis,LunLi2016} with emphasis on utilizing simulation techniques to justify the structurally resonant passivity of the gondola as it pertains to pointing performance.

In order provide diffraction limited imaging at the telescope focal plane with a 0.02$^{\prime\prime}$ effective image stability, \superbit architecturally refines its pointing precision through three successive pointing and stabilization regimes: coarse target acquisition to within < $0.5^\prime$, fine telescope stabilization at the 0.5-1$^{\prime\prime}$ level (1-$\sigma$) , and image stabilization at $0.02^{\prime\prime}$ (1-$\sigma$) .
Firstly, targets are acquired for accessible regions on the sky by slewing the three gimbal frames to coordinates corresponding to the desired right ascension (RA) and declination (Dec) of the target. 
Gondola yaw is coarsely achieved via a large 1000 N$\cdot$m$\cdot$s reaction wheel, which is driven by a three-phase torque-controlled frameless motor on the base of the outer frame, with inertial and absolute feedback provided by fiber optic rate gyroscopes and a three-axis magnetometer.
To prevent reaction wheel saturation due to persistent disturbances, excess angular momentum is dumped through the gondola and along the flight train to the balloon via a speed-controlled two-phase pivot stepper motor, which provides the necessary gondola-external torque by manipulating flight train twist. 
Interestingly, it can be shown\cite{RomualdezThesis18} that this approach to yaw stability not only provides yaw control, but also damps compound pendulation effects from the balloon, a result that allows for suppression of potentially image-degrading resonant modes.
Along the roll and pitch axes, the middle and inner frames, respectively, are each actuated about their respective frames via pairs of high-inductance three-phase frameless motors equipped with flexure bearings, to prevent non-linear static friction effects, as well as high resolution absolute optical encoders for relative gimbal position feedback. 
In order to access the full elevation range of the telescope, the pitch axis of the inner frame is co-actuated by a pair of two-phase stepper motors through a high resolution 100:1 gear box, which together coarsely acquires the necessary pitch gimbal position to with <1$^\prime$. 

\begin{table}
\caption{\superbit attitude sensor characteristics \cite{RomualdezThesis18}}
\label{table:sensor_char}
\begin{tabular}{*{1}{l}*{3}{c}r}
	\toprule
	Sensor Description & Readout Frequency (Hz) & Resolution & Noise Figure	\\
	\midrule
	Fibre optic rate gyroscope & 1000\textsuperscript{a} & $4.768\cdot 10^{-4}$ deg/s & $2.2\cdot 10^{-4}$ deg/(s$\cdot\sqrt{\mbox{Hz}}$)\\
	Absolute optical encoder & 100 & $5.49\cdot 10^{-3}$ deg & -- \\
	3-axis magnetometer & 20 & $6.7\cdot 10^{-5}$ Gs & $2.0\cdot 10^{-4}$ Gs\\
	Coarse elevation stepper & 10 & $9.374\cdot 10^{-3}$ deg & --\\
	Bore star camera & 20 & $0.23^{\prime\prime}$ centroids & $5.75\cdot {10^{-4}}^{\prime\prime}$/s\textsuperscript{b}\\
	Roll star camera & 20 & $0.46^{\prime\prime}$ centroids & $5.75\cdot {10^{-4}}^{\prime\prime}$/s\textsuperscript{b}\\
	Focal plane star camera & 60 & 0.025$^{\prime\prime}$ & $5.77\cdot {10^{-5}}^{\prime\prime}/\mbox{s}$\textsuperscript{b}\\
	\bottomrule
\end{tabular}
\\
\small{\textsuperscript{a} Asynchronous serial ($\pm 5\%$) remapped to synchronous 1000 Hz via Akima interpolation \cite{RefWorks:92}}
\\
\small{\textsuperscript{b} Sky equivalent read noise}
\end{table}
\begin{table}
\caption{SuperBIT actuator characteristics \cite{RomualdezThesis18}}
\label{table:actuator_char}
\begin{tabular}{*{1}{l}*{2}{c}r}
	\toprule
	Actuator Description & Control Input & Characteristics\\
	\midrule
	Reaction wheel - frameless motor & 16-bit analog & 15 N$\cdot$m max.; 3600 lines/rev incremental encoder\\
	Pitch/roll - frameless motor $\times$ 4& 8-bit PWM\textsuperscript{a} & 5.0 N$\cdot$m max. torque; 3-phase Hall sensor feedback\\
	Pivot - stepper motor & step/direction & 0.018 deg/step\textsuperscript{b}; 2-256 \si\micro step/step; 0.44 N$\cdot$m hold \\
	Pitch - stepper motor $\times$ 2 & step/direction & 0.15 deg/step\textsuperscript{c}; 16 \si\micro step/step; 0.51 N$\cdot$m hold\\
	Piezo-electric tip/tilt & 16-bit analog & $0.027^{\prime\prime}$ resolution per axis; 45 Hz bandwidth\\
	\bottomrule
\end{tabular}
\\
\small{\textsuperscript{a} Pulse-Width Modulation}\\
\small{\textsuperscript{b} 1.8 deg/step motor through a 100:1 gear reducer}\\
\small{\textsuperscript{c} 1.8 deg/step motor through a 12:1 gear reducer}
\end{table}
After coarse target acquisition, a pair of orthogonal star cameras provide both absolute pointing information through so-called \textit{lost-in-space} routines using the pattern recognition routines developed for \texttt{astrometry.net}\cite{dustin} as well as differential feedback for high bandwidth control telescope stabilization at the sub-arcsecond level through high resolution centroiding. 
As shown in Table \ref{table:actuator_char}, the star cameras can centroid to sub-arcsecond accuracies, and are sensitive to stars of at least Mag 9, which gives on average 4 stars per field of view \cite{Zotov}; this is essential for the lost-in-space pattern matching algorithm \cite{dustin} when knowledge of the telescope attitude and camera pixel scale are poorly constrained.
Although lost-in-space solutions provide a full pointing solution every 2 seconds, higher rate differential feedback is necessary to mitigate mid-frequency perturbations in the balloon-borne environment. 
In this differential mode, the star cameras provide star centroiding of the brightest stars in their fields-of-view in order provide frameless motor feedback with the corrections necessary to maintain telescope stability whilst serving as an absolute reference to continually correct for drifting fiber-optic rate gyroscope biases. 
Sub-pixel star centroiding is possible based on a deliberate star camera CCD and lens selection for \superbit (\ref{table:sensor_char}), which is specified to oversample the stars' PSFs on the the star cameras' focal planes sufficiently to obtain approximately 1/10th pixel centroiding precision.
With the coarse stepper motors locked during fine telescope stabilization, the combination of rate gyroscopes and star cameras provide feedback for frameless motors on all three gimbaled frames to provide the necessary three-axis stability for the telescope. 

Once stabilized to <1$^{\prime\prime}$ (1-$\sigma$) , the telescope back-end optics further reduce the focal plane disturbances to 0.02$^{\prime\prime}$ (1-$\sigma$) using a piezo-actuated tip/tilt fold mirror along the optical path of the telescope between the corrective lenses and the science camera CCD. 
In order to provide feedback for the tip-tilt actuator, a portion of the science focal plane is picked-off and redirected to focal plane star camera that provides high rate differential feedback through sub-pixel centroiding.
This centroid information is improved by incorporating rate gyroscope data such that the effective correction bandwidth of the image stabilizer is limited by the inertia of the fold mirror used for tip-tilt corrections as well as the relative power of the piezo-electric controller.
Given the potential for thermal misalignments along the telescope boresight, the focal plane star camera is equipped with an independent linear actuator to decouple its focus control from the telescope secondary focus position. 
Systematic calibrations between the telescope stabilizing star cameras and the science camera focal plane are determined through similar lost-in-space routines performed on the science camera focal plane. 
Lastly, in-flight telescope re-alignment is made possible via secondary mirror actuators that provide tip/tilt corrections that correct for potential optical misalignments suffered due to thermal effects or launch shocks.

\subsection{Software \& Control Architecture}

Effective implementation of control software is the backbone that enables mechanical hardware on \superbit to achieve pointing and image stability at the $0.02^{\prime\prime}$ level.
The theoretical development of control algorithms and use of hard real-time operating systems with effective timing and synchronization allow for on-demand, robust, and repeatable pointing performance within the balloon-borne environment, where the mathematical basis for attitude determination, stability, and control can be rigorously shown to provide asymptotic stability under any number of pointing regimes. 
A brief description of some of the important approaches to this kind of real-time control is provided here with reference to a more in-depth treatment and justification for particular software architecture available in related works.\cite{RomualdezThesis18, Romualdez16}

\begin{figure}[h!]
\centering
\includegraphics[width=0.9\textwidth,page=1]{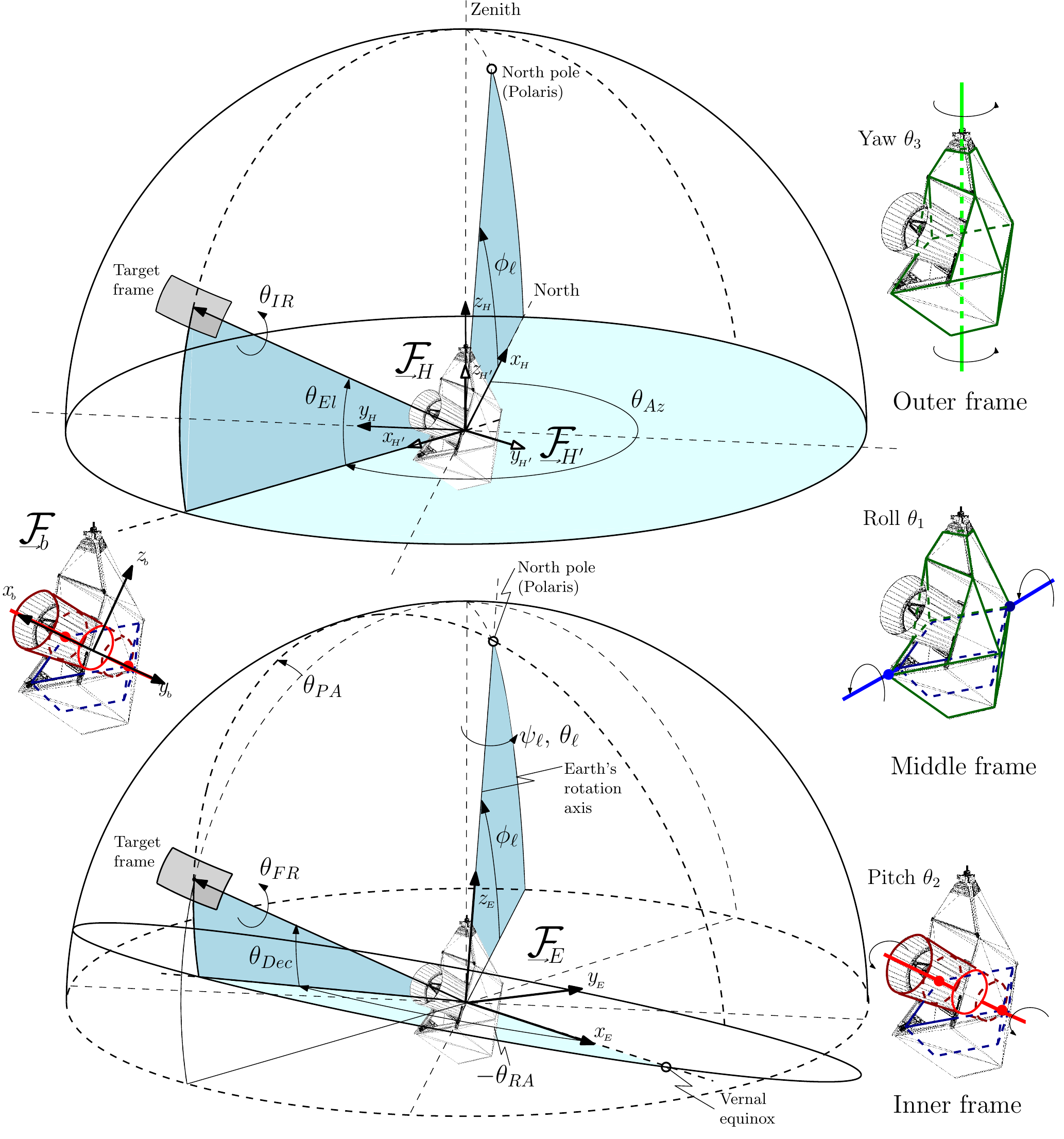}
\caption{Relevant coordinate systems for scientific balloon-borne payload with reference to the \superbit gondola physical architecture\cite{RomualdezThesis18}; local horizontal (top) and equatorial (bottom) coordinates are given for an arbitrary target and related to the \superbit gimbal coordinate system}
\label{fig:coordinate_systems}
\end{figure}

With regards to coarse target acquisition, large slews are prescribed a trapezoidal speed profile such that a maximum permissible acceleration and angular speed can be effectively set.
The RA and Dec of a given target on the sky are projected onto the three gimbal axes in order to determine the necessary gondola yaw, roll, and pitch that corresponds to the given target.
Since there is one more extra degree-of-freedom than is typically specified for a given celestial target, the roll of the middle frame is free to be set at any angle within its range that maximizes integration time on the sky, such that science observations are limited by the $\pm 6^\circ$ constraint of the middle frame; in the worst case, this allows for $\sim 1$ hour integration times integration time for a given field rotation before the gimbals need to reset at a different field rotation for a given target. 
Towards the end of a gimbal slew, control effort smoothly transitions to a fully-coupled attitude controller in preparation for telescope stabilization.

During pointing control and stabilization, in particular for the fine telescope stabilization phase, all pointing sensors, including star camera measurements, rate gyroscope measurement, encoder readings, etc., are optimally integrated into a single pointing solution using the canonical extended Kalman filter (EKF) for attitude determination, the development of which utilized perturbation methods for effective and consistent state estimation \cite{BarfootGeneral}. 
Additionally, the EKF continuously makes use of fused sensor data to estimate biases in the rate gyroscopes, which suffer detrimentally from large thermally-depended steady-state errors, while providing modes for estimating calibration matrices that account for potential axes or sensor misalignments as well as outer frame imbalances.
Based on the fused pointing solution, high rate feedback for the \superbit actuators is generated, where the pointing error with respect to the desired target is projected onto the gimbaled axes and used to generate a globally asymptotically stable control effort that robustly stabilizes the telescope to the sub-arcsecond level.
An in-depth discussion and a rigorous background for the state estimation and gondola control techniques developed for general scientific balloon-borne payloads with an emphasis on \superbit are given in a related work \cite{RomualdezThesis18}.

As previously mentioned, feedback for further image stabilization is provided by the focal plane star camera, which provides feedback to the tip-tilt actuator for diffraction-limited imaging on the \superbit focal plane at the $0.02^{\prime\prime}$ level (1-$\sigma$) . 
A single-state EKF implementation is used to fuse centroiding measurement information from the focal plane star camera with high resolution, low-noise rate gyroscopes, where robust timing of star camera exposures is provided in software via real-time implementation. 
Since both telescope and image stabilization are actively and concurrently stabilizing the focal plane in the \superbit control architecture, it is interesting to note that controller gain and attenuation trade-offs need to be considered in order to make effective use of the tip-tilt actuator bandwidth; in this way, fine tuning of both control loops must be done concurrently during initial telescope operations in flight in order to achieve robust and on-demand image stability for observations.
To this end, a reliable remote teleoperations system was developed for \superbit to facilitate the both the commanding and telemetry to and from the telescope payload as well as to downlink compressed science data for initial data validation during flight.

\section{ENGINEERING TEST FLIGHTS}
\label{sec:testflights}

\subsection{Timmins 2015 BIT Flight}
\begin{figure}
\centering
\includegraphics[width=1.0\textwidth]{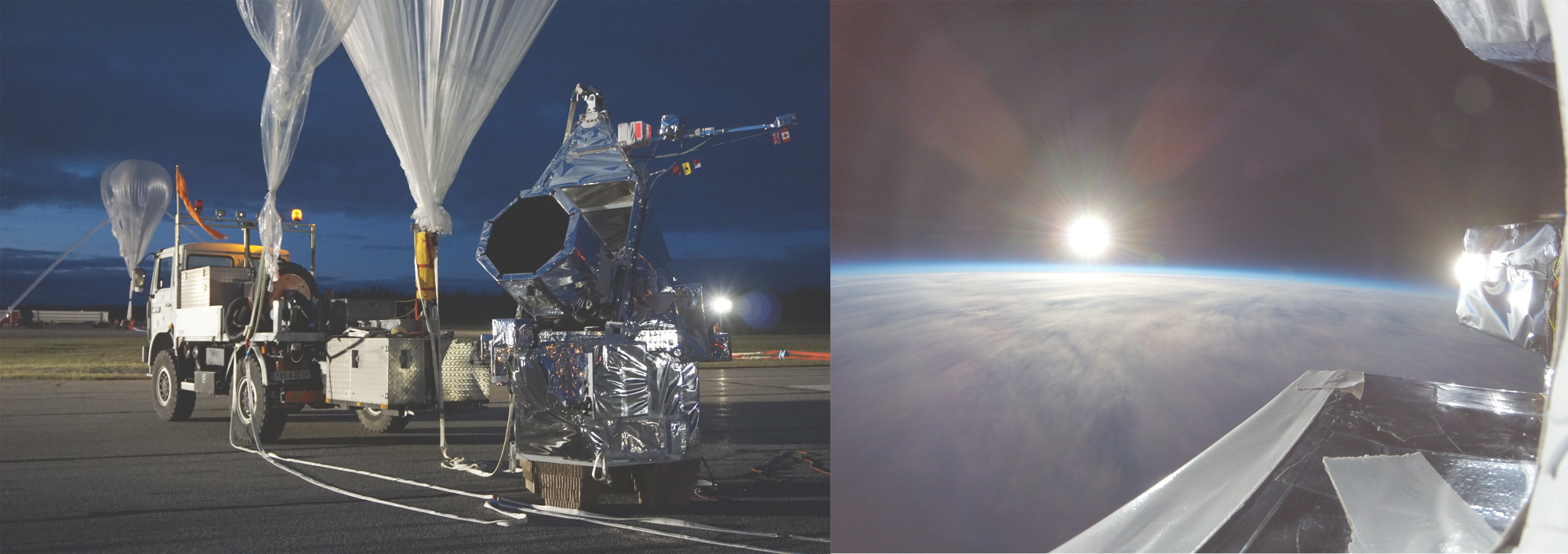}
\caption{The BIT 2015 Timmins campaign and flight; (left) the BIT gondola on the launch pad hours before launch on September 18, 2015; (right) footage from flight after pointing operations, approximately 1 hour before balloon termination and descent via parachute on the morning of September 19, 2015}
\label{fig:2015_photos}
\end{figure}

The inaugural engineering flight of the Balloon-borne Imaging Testbed (aka. BIT: the predecessor to \superbit) took place over a single night from September 18--19, 2015.
It was launched from the Timmins Stratospheric Balloon Base in Ontario, Canada, with facilities provided by the Canadian Space Agency (CSA) and launch hardware provided by the centre national d'\'etudes spatiales (CNES). 
Remote communications to and from the BIT gondola at 1 Mbps were made possible via a high-gain line-of-sight link, allowing for human-in-the-loop teleoperation throughout the flight. 
The BIT payload remained at 36 km altitude for 6.5 hours, descended via parachute in Northern Quebec, and was safely recovered with minimal damage.

The primary goal of this flight was to test the engineering capabilities of the BIT pointing system, in particular telescope stabilization via absolute and differential star camera feedback as well as fine image stabilization via back-end optics; thus, the constraints on coarse target acquisition were not as stringent, where star fields were used to calibrate star camera pointing sensors and focal plane image stabilization feedback. 
During development, BIT was designed as a proof-of-concept for \superbit as well as a way to characterize potential system performance within the harsh dynamic and thermal environment of the upper atmosphere.
Over several test tracking and stabilization runs during the flight, the BIT pointing instrumentation successfully stabilized the telescope to within $0.68^{\prime\prime}$ ($1\sigma$) for integration periods as long as 1.4 hours, which was verified both during flight and post-flight based on the star camera centroid dispersion during a given tracking run (e.g. Fig. \ref{fig:2016_pointing_stab}). \cite{Romualdez16}
Regarding attitude determination, estimator consistency agreed quite with the estimated pointing errors on the sky, where pointing performance was purely limited by integration of rate gyroscopes between star camera measurements.

For a few of these successful telescope tracking runs, a bright star was manually placed on the telescope focal plane star camera, through small relative corrections to the telescope attitude with respect to the sky, in order to perform testing and calibration for image stabilzation routines.
Although diffraction-limited imaging was not observed during this flight, telescope alignment was sufficient to demonstrate a 1-$\sigma$ image stability of $0.12^{\prime\prime}$ on the telescope focal plane, which, after post-flight analysis, was limited by the bandwidth of the original piezo-electric tip-tilt actuator.\cite{Romualdez16}
Furthermore, science camera images of large star fields were taken over 10-20 minutes long integration periods to assess the beam quality of the telescope post-launch, which clearly showed that either telescope alignment pre-flight was insufficient or that alignment had suffered severely from launch shocks. 

Despite clear improvements to telescope optical alignment and image stabilization hardware, the 2015 Timmins flight successfully demonstrated the ability to achieve sub-arcsecond pointing and image stability from a balloon-borne platform, which had previously not been demonstrated at that level of precision, duration, and repeatability. \cite{RefWorks:95} 
Following the BIT 2015 flight, the substantiated potential for refurbishing the gondola for an SPB platform (aka. \superbit) had been realized, where the general foundation and methodologies for precise pointing had been shown as a proof-of-concept during flight.
To that end, refurbishments were made post gondola recovery to increase the overall bandwidth of the image stabilization stage while improving the ability of the BIT system to accurately acquire targets of interest, which could be driven by science objectives rather than engineering demonstrations. 
Furthermore, a more rigorous approach to telescope alignment pre-flight was developed in order to better assess how telescope performance would change post-launch and throughout the flight due to changes in mechanical and thermal loading. 

\subsection{Palestine 2016 \superbit Flight}
\begin{figure}
\centering
\includegraphics[width=1.0\textwidth]{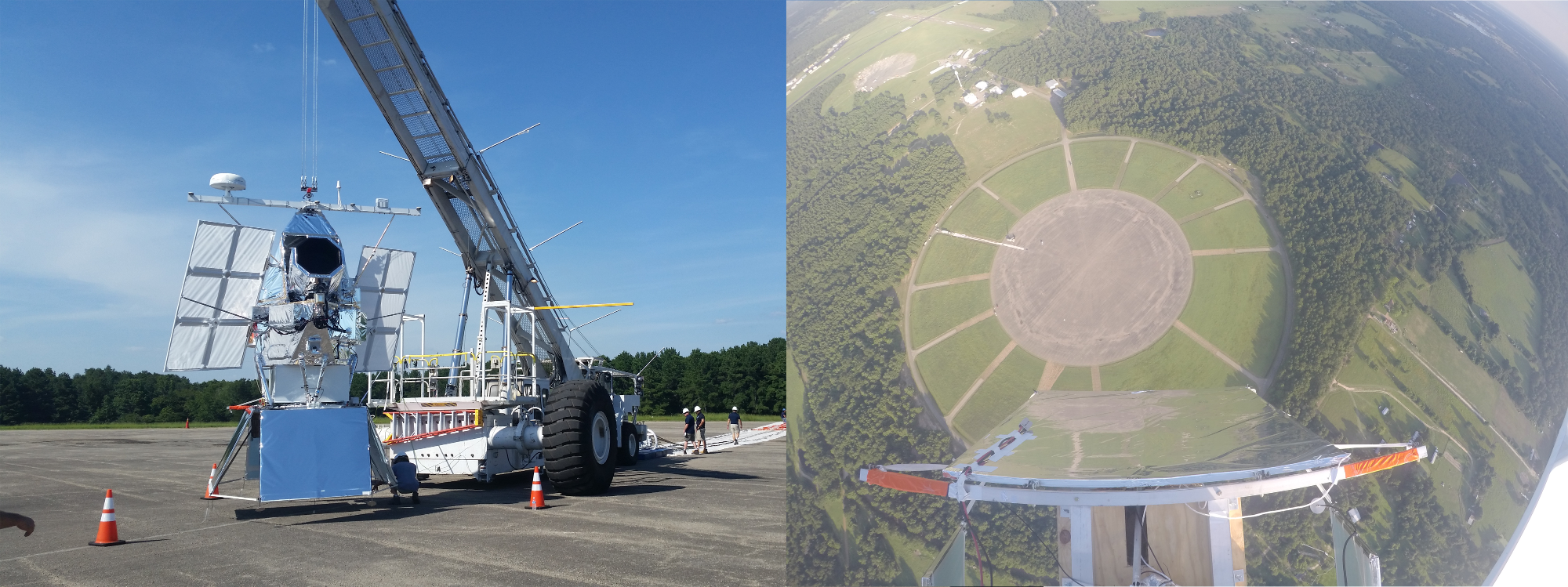}
\caption{The \superbit 2016 Palestine campaign and flight; (left) the \superbit gondola on the launch pad hours before launch on June 30, 2016; communications antennas and mass-model solar panels were used to accurately emulate SPB configuration; (right) footage taken from a downward facing GoPro camera on the \superbit gondola seconds after launch from the CSBF launch pad}
\label{fig:2016_photos}
\end{figure}

Launched from the Columbia Scientific Balloon Facility (CSBF-NASA) located in Palestine, TX for a single night flight from June 30--July 1, 2016, the \superbit 2016 flight was a direct follow-up based on the engineering results from the 2015 Timmins flight.
Given the prospects for CSBF to provide SPB capabilities for long-duration mid-latitude flights, one of the goals of this flight was to demonstrate the ability to operate and calibrate the pointing systems during the flight using communications hardware and protocols similar to what would be used for a SPB flight, which includes a suite of line-of-sight and over-the-horizon telemetry and commanding links at various bandwidths. 
Based on the results from the previous flight, further the engineering goals for the \superbit 2016 flight included a more accurate and robust target acquisition stage for observing particular regions of astronomical interest, reconfirmation of telescope pointing stability at the $<1^{\prime\prime}$ level, and improved image stability with redesigned, higher bandwidth piezo-electric tip-tilt hardware. 
Furthermore, improved techniques for aligning the \superbit telescope pre-flight were used with telescope support structure modifications to help increase optical alignment robustness to mechanical shock from launch. 

\begin{figure}
\centering
\includegraphics[width=0.95\textwidth]{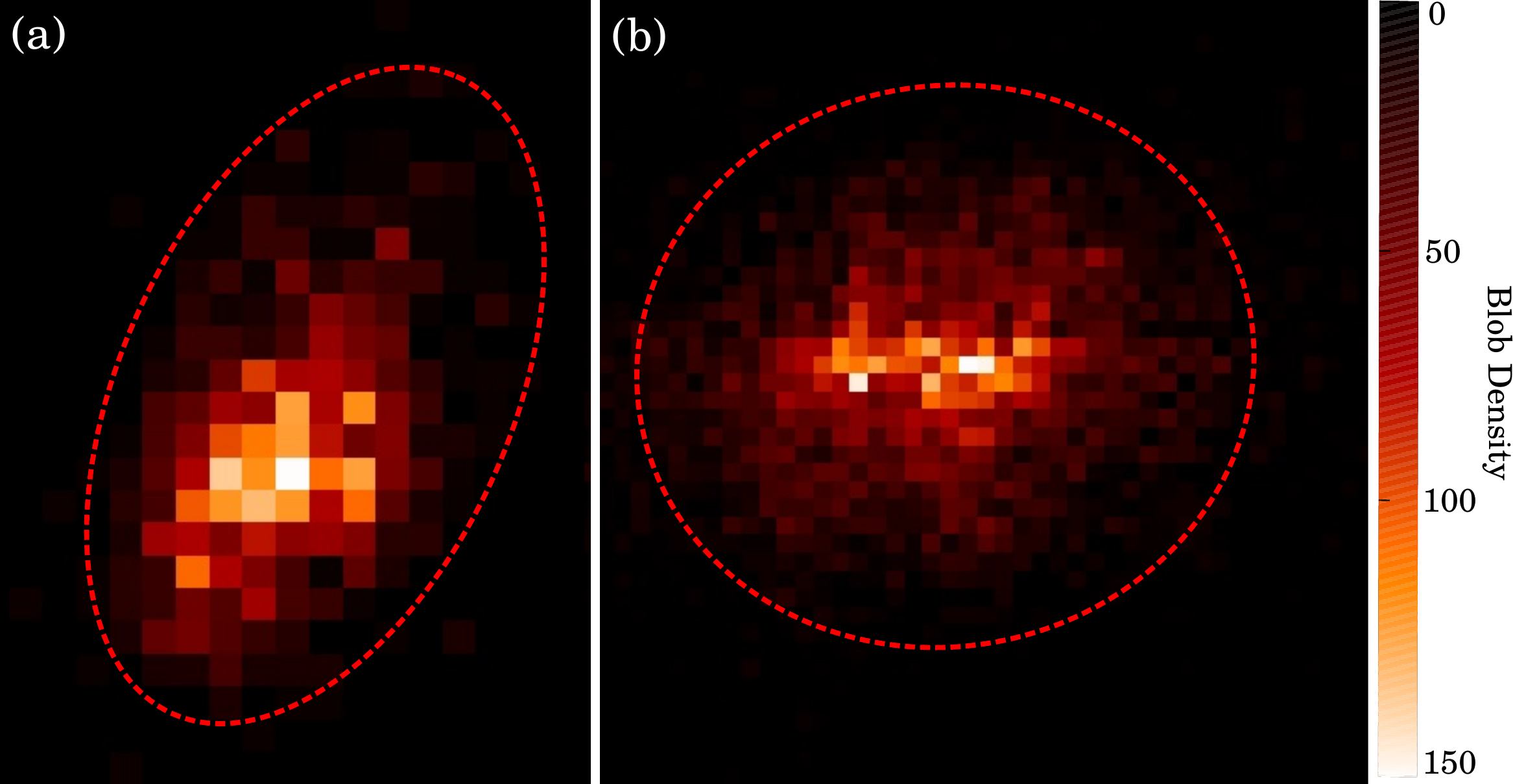}
\caption{Fine telescope and image stabilization during the \superbit 2016 Palestine flight illustrated through star camera centroid distributions during a typical observing run (10-50 minutes); (left) fine telescope stabilization is shown on one of the telescope star cameras with 0.43$^{\prime\prime}$ binning; (right) image stabilization is shown on the focal plane star camera with 0.023$^{\prime\prime}$ binning; the red dashed ellipses for each represent the three-sigma distribution bounds for the centroids}
\label{fig:2016_pointing_stab}
\end{figure}

During flight, a number of targets were successfully acquired to within $<1^{\prime}$ pointing accuracy with the ability to automatically and reliably drop and tracking star on the focal plane star camera on-demand for a given target of interest about which the telescope was stabilized.
On average, a similar rate-gyroscope-limited three-axis stabilized pointing performance was observed at the $0.7^{\prime\prime}$ level (1-$\sigma$) with increased robustness due to a higher star camera centroid rate from improved firmware and readout. 
As with the previous flight, pointing performance was verified both in-flight and post-flight through the mapping of centroid distributions over a particular tracking run, which, for this flight, took place over $\sim 0.5$--$1$ hour periods. 
It was observed that pointing performance was degraded at time in the strictly azimuthal direction for 1-5 minute periods at a time, which could be attributed to the increased inertia and structural flexibility induced by the simulated solar panels; however, as shown in Fig. \ref{fig:2016_pointing_stab}, centroid distribution on the focal plane star camera remained unaffected overall due to effective image stabilization.
A more in-depth analysis and discussion on the results of telescope stabilization for this particular flight is provided in related work \cite{RomualdezThesis18}. 

\begin{figure}
\centering
\includegraphics[width=1.0\textwidth]{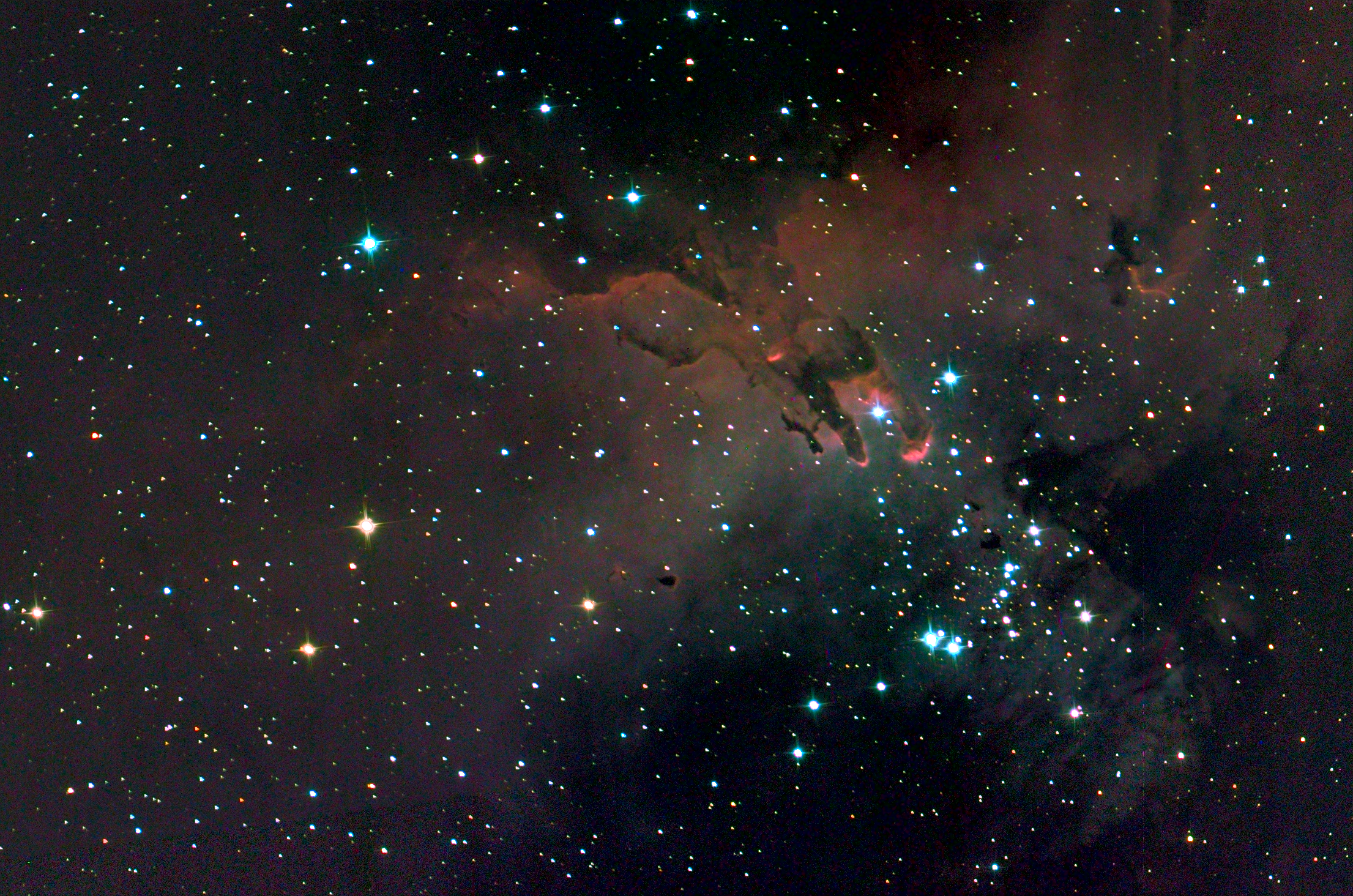}
\caption{One of the ``glamour shots'' taken by the \superbit instrument during the Palestine 2016 flight\cite{RomualdezThesis18}/; this image of the Eagle Nebula is a composition of several 1-3 minute integrations in several observing bands from near-IR to near-UV; the total observing time required to acquire the complete composite image was 17 minutes from the beginning of the first to the end of the last exposure, where the size of the image represents the 0.5$^\circ$ field-of-view of the telescope}
\label{fig:eagle_nebula}
\end{figure}

The ability to place a suitable tracking star on the focal plane star camera allowed for more opportunities to assess the performance of the image stabilization hardware and routines used for the \superbit 2016 flight, where each telescope tracking run was motivated either by science camera and telescope optics characterization or by science-related objectives.
Due to improved image stabilization hardware, the 1-$\sigma$ image stability on the telescope focal plane was repeatedly demonstrated at the $90$ milli-arcsecond level over 20 minutes, which, from the post-flight results, appeared to now be limited by the bandwidth of the focal plane camera centroid readout rate as well as the noise level on rate gyroscope feedback.
Furthermore, the beam quality observed during flight and extracted from science images post-flight suggest not only that launch shocks had exceeded the permissible dynamic load on the telescope optics to maintain alignment, but that methodologies to align and validate telescope optics pre-flight may have also been inadequate.
In short, these results emphasize the need to remotely realign the telescope post-launch, a capability that has since been implemented post-recovery of the \superbit payload in 2016. 

Despite the reduced beam quality, it is clear from the imaging and engineering results that the capability to obtain sub-arcsecond precision imaging from a balloon-borne payload is not only feasible, but is robust and repeatable despite the harsh dynamic sub-orbital environment, as demonstrated from the \superbit 2016 flight.
A single image shown in Fig. \ref{fig:eagle_nebula} illustrates the ability of the \superbit instrument to obtain high-resolution imaging from the stratosphere over a wide field in a number of optical bands from near-IR to near-UV.
Overall, it is reasonable to assert that the positive results from this flight highlight the potential for the \superbit instrument to generate high science return during a prospective SPB flight, which, at the time of this writing, will soon be made available.

\subsection{Palestine 2018 \superbit Flight}
To further demonstrate and improved upon \superbit instrumentation for SPB, \superbit will launch its third overnight engineering flight from Palestine, TX with CSBF-NASA in June of 2018. 
This will be the final flight with the original BIT 2015 telescope optics as the existing optical hardware has now approached end-of-life.
In order to address the clear discrepancy between the expected 20 milliarcsecond pointing requirement and the observed image stability during the Palestine 2016 flight, state-of-the-art fibre-optic rate gyroscopes have been implemented alongside a high-speed, highly sensitive focal plane camera in order to increase the responsivity and bandwidth of the image stabilization stage, both of which will be fully tested on potential science targets during the Palestine 2018 flight.
Similarly, the thermal performance of the telescope optics and the subsequent effects on beam quality will be fully assessed during the flight through observation of a number of calibration targets under various thermal conditions.
From a post-flight analysis perspective, particular science targets have been chosen to assess the viability of a newly developed data analysis pipeline for \superbit, which will be invaluable for eventual use on data produced from the SPB platform.
In short, these and other similar engineering and science goals for the Palestine 2018 will mainly serve as probes to better inform methodologies not only for the SPB flight of \superbit, but for other instruments hoping to benefit from visible-to-near-UV imaging from the stratosphere.

\section{SCIENCE FLIGHTS}
\label{sec:operationflights}

\subsection{Super Pressure Balloon (SPB) Flight}
\superbit is scheduled for a 100 day super presure balloon flight from Wanaka, New Zealand by 2020. 
From an engineering perspective, the improvements that are essential to maintaining \superbit's optimal performance over three months center on increased autonomy, downlink economy, and long-term robustness existing systems, given the increased operational lifespan of the gondola over several months and the intermittent communication with flight systems over that time. 
More specifically, the \superbit gondola must self-regulate power and thermal subsystems over diurnal cycles at mid-latitudes, while preventing damage to sensitive optical components while charging batteries and calibrating optics during the day and mitigating a fairly large temporal cross-section for cosmic-ray upsets in computer and memory subsystems.
Given the variations in temperature and altitude experienced on a daily basis, \superbit on a SPB platform must have the ability to be remotely aligned effectively and efficiently as to maximize observation time during the night.
Furthermore, effective methods for downlinking compressed science data during flight is essential to assessing the quality of the data during the flight as well as to plan daily/weekly observations based on preliminary results from mid-flight data analyses; in fact, at the time of this writing, methods are currently being explored for recovering the full data payload in the event that the payload is irrecoverable.

For the prospective SPB flight in 2020, the \superbit gondola will be retro-fitted with a custom f/11 telescope designed with increased robustness to launch shocks and thermal gradients as well as improved hardware for mid-flight alignment and beam quality assessment.
The science-drive goals of the \superbit SPB flight include obtaining high-precision weak and strong gravitational lensing measurements of $\sim$200 galaxy clusters at redshifts $0.1<z<0.5$, and their attachment to the cosmic web.
\superbit's cluster sample is intentionally compiled from a variety of X-ray, Sunyaev-Zel'dovich and optical/NIR imaging surveys, where cross-calibration of cluster mass measurements by a single, high-precision instrument will resolve current discrepancies between cosmological tests, which may stem from foregrounds in the CMB, mis-calibration of X-ray telescopes, or previously unknown physics.
Furthermore, the increasing number and mass of clusters towards low redshift, reflecting the gradual formation of cosmic structure, is a sensitive probe of the nature of Dark Energy.

\subsection{Facility Class Instrument}
\superbit is a step toward a future platform that can accommodate facility-class instruments requiring sub-arcsecond stability for annual suborbital flights of up to 100 days in the early 2020's. Such a mission will be able to carry lightweight mirrors between 1 and 2 meters. 
In combination with a giga-pixel class focal plane (compared to 32MP for \superbit), these systems will exceed the imaging capability of HST and will play a highly flexible role among the fixed concert of Euclid, WFIRST and LSST.
Below, we summarize some of the broad advantages of stratospheric balloon-based observations that motivate long-term development of a flexible observing platform:
\paragraph*{UV photometry:} Strong UV absorption by the atmosphere makes wide-field near-UV imaging from the ground restrictively inefficient and time consuming. However, this data is critical for the accurate determination of photometric redshifts that are needed, for instance, in dark energy studies. A wide-field UV imaging survey would therefore benefit both Euclid and WFIRST. As illustrated in Fig.~\ref{fig:balloontrans}, observations from stratospheric altitudes suffer significantly less UV absorption and atmospheric background than from the ground. A meter-class telescope, even coupled to a modest camera, has similar survey speed to LSST in its bluest band, but at a much higher resolution while extending to 300\,nm wavelength, allowing the terrestrial telescope to focus on the visible and near-IR.
\paragraph*{Persistent, sub-arcsecond imaging:} 
As demonstrated by the ongoing, $8\times$ over-subscription of HST, even 25 years after its launch, many branches of astronomy exhibit a persistent need for high resolution UV/optical/IR observations. For example, observations of low-surface-brightness galaxies (and satellites of our own Milky Way) and a census of the stars within the Galaxy and nearby galaxies (galactic archaeology) would all greatly benefit from the high resolution of an optical-UV balloon-borne telescope. Exoplanet searches, such as microlensing studies of stars toward the Galactic bulge, also benefit from high-resolution imaging, over a long time baseline of observations. A balloon-borne microlensing program covering, for example, some of the proposed WFIRST microlensing fields, would not only provide added science value to that mission by stretching this baseline, but also be immune to the weather-induced losses of observations that can interrupt the critical cadence needed for these observations. 
\paragraph{Technological development:} The path to developing new sensors, detectors, and instruments for space missions is difficult and expensive. Few opportunities exist to fly new technologies in a space-like environment at a reasonable cost. A facility class instrument flying regularly on the SPB platform with a relatively low launch cost, will provide a technology testbed to help ensure space qualification of components and instruments that will become the backbone of future flagship, probe, and Explorer-class space missions.

\section{CONCLUSIONS}
\label{sec:conclusions}
Progress in many branches of cosmology and astrophysics currently relies on space-quality observations. From below the Earth's turbulent atmosphere, astronomy is limited by the blurring or blending of adjacent sources, and the misidentification of Milky Way stars with distant galaxies. Recent advances in multi-conjugate adaptive optics have helped some of the science cases that require only a narrow field-of-view. However, rising above the atmosphere will remain the only solution for science cases that require accurate knowledge of the PSF, or a wide field of view. 

We report the successful demonstrations of a balloon-borne telescope \superbit, which has been shown to have robust, repeatable, and long-during wide-field sub-arcsecond imaging capabilities at stratospheric altitudes above 99\% of the Earth's atmosphere.
At a cost less than 1\% the cost of an equivalent satellite, but in conjunction with advances in `superpressure' balloon technology that recently extended flight duration from $\sim3$ days to $\sim3$ months, \superbit offers simply transformative opportunities, where the inaugural \superbit 100-day SPB flight is scheduled for the spring of 2020.

This is merely a first step towards an ambitious but achievable goal of facility-class 1--2 m telescopes providing hundreds of days of near space-quality imaging and spectroscopy. 
These missions will be a cutting edge but flexible facility, able to rapidly adapt to the most interesting science areas throughout the 2020s. 
Of particular interest will be the interplay of these SPB missions with large ground and space-based missions, such as Euclid, LSST, and WFIRST. 
In addition to quickly following up currently unimagined science goals, \superbit's UV and high resolution complementarity offers a unique potential to enhance science return even in key research areas including exoplanets, dark energy, and dark matter science.

\acknowledgments
\superbit is supported in Canada, via the Natural Sciences and Engineering Research Council (NSERC), in the USA via NASA award NNX16AF65G, and in the UK via the Royal Society and Durham University. Part of the research was carried out at the Jet Propulsion Laboratory (JPL), California Institute of Technology (Caltech), under contract with NASA.

\bibliography{report} 
\bibliographystyle{spiebib} 

\end{document}